\documentclass[12pt]{article}

\usepackage{fullpage}
\usepackage{amssymb}
\usepackage{bbm}
\usepackage{color}
\usepackage{ulem}

\DeclareFontFamily{OT1}{rsfs10}{}
\DeclareFontShape{OT1}{rsfs10}{m}{n}{ <-> rsfs10 }{}
\DeclareMathAlphabet{\mathscript}{OT1}{rsfs10}{m}{n}

\def\gsim{ \lower .75ex \hbox{$\sim$} \llap{\raise .27ex \hbox{$>$}} }
\def\lsim{ \lower .75ex \hbox{$\sim$} \llap{\raise .27ex \hbox{$<$}} }
\def\be{\begin{equation}}
\def\ee{\end{equation}}
\def\bea{\begin{eqnarray}}
\def\eea{\end{eqnarray}}

\newcommand{\ns}{\normalsize}

\usepackage{latexsym,amsmath,amssymb,epsfig}

\usepackage{graphicx}
\usepackage{graphicx,subfigure}
\usepackage{epstopdf}
\usepackage[body={17.5cm, 21cm},right=2cm]{geometry}
\usepackage{amssymb}
\usepackage{amsmath}
\usepackage{psfrag}
\usepackage{epsfig}
\usepackage{cancel}
 \allowdisplaybreaks[4]

\usepackage[all]{xy}

\begin{document}

\begin{titlepage}

\vspace{-5cm}

\title{
  \hfill{\ns }  \\[1em]
   {\LARGE Scale Invariance via a Phase of Slow Expansion}
\\[1em] }
\author{
   Austin Joyce and Justin Khoury
     \\[0.5em]
{\ns  Center for Particle Cosmology, Department of Physics and Astronomy,} \\[-0.4cm]
{\ns  University of Pennsylvania, Philadelphia, PA 19104}\\[0.3cm]
}

\date{}

\maketitle

\begin{abstract}
We consider a cosmological scenario in which a scale-invariant spectrum of curvature perturbations is generated by a rapidly-evolving equation of state on a slowly expanding background.
This scenario generalizes the ``adiabatic ekpyrotic" mechanism proposed recently in arXiv:0910.2230. Whereas the original proposal assumed a slowly contracting background, the
present work shows that the mechanism works equally well on an expanding background. This greatly expands the realm of broader cosmological scenarios in which this mechanism can be embedded. We present a phase space analysis and show that both the expanding and contracting versions of the scenario are dynamical attractors, with the expanding branch having a broader basin of attraction. In both cases, a finite range of scale invariant modes can be generated within the regime of validity of perturbation theory. 
\end{abstract}

\end{titlepage}

\section{Introduction}

There is mounting observational evidence that the large scale structure originated from a nearly scale invariant and nearly Gaussian spectrum of
primordial density perturbations. While these statistical properties are consistent with the simplest inflationary models~\cite{inf}, a critical
question for early universe cosmology is whether inflation is unique in making these predictions. This has motivated the quest for
alternative scenarios, from the pre-big bang scenario~\cite{Gasperini:1992em}, to string gas cosmology~\cite{Nayeri:2005ck}$-$\cite{Battefeld:2005av},
to ekpyrotic theory~\cite{Khoury:2001wf}$-$\cite{Lehners:2010fy}. (See~\cite{Lehners:2008vx,Lehners:2010fy} for reviews of ekpyrotic cosmology.)

A zeroth-order benchmark for a successful theory of the early universe is to explain the observed homogeneity, isotropy and flatness of our universe.
There are only two known cosmological phases that make the universe increasingly homogeneous, isotropic and flat. The first is of course inflation,
characterized by accelerated expansion after the big bang. This requires a scalar field (or fluid) with negative equation of state $w < -1/3$.
A second possibility is ekpyrosis, a phase of slow contraction before the big bang~\cite{Gratton:2003pe,Creminelli:2004jg,Erickson:2003zm}. This corresponds to a scalar field
with stiff equation of state $w \gg 1$. The smoothing power of ekpyrotic contraction was recently confirmed using numerical relativity simulations~\cite{Garfinkle:2008ei}.

Another benchmark is to generate a nearly scale invariant and Gaussian primordial spectrum. 
But even this is not sufficient. Inflation not only generates perturbations with the desired properties,
but it does so within a cosmological background that is a dynamical attractor. Indeed, on super-horizon scales the curvature perturbation on
uniform-density slices~\cite{zeta1}$-$\cite{separateuniverse}, $\zeta \approx \delta a/a$, measures differences in the expansion history of
distant Hubble patches~\cite{separateuniverse}. Since $\zeta\rightarrow {\rm const.}$ at long wavelengths in (single-field) inflation, 
the background is an attractor~\cite{weinbergzeta}.

Achieving both scale invariance and dynamical attraction in alternative scenarios has proven challenging.
The mode function equation for $\zeta$ in a contracting, matter-dominated universe takes an identical form as in inflation~\cite{dust,latham,robertmatter};
but $\zeta$ grows outside the horizon in this case, indicating an instability. This is not surprising, since an anisotropic stress component, for instance,
will blueshift faster than dust.  Similarly, contracting mechanisms that rely on a time-dependent sound speed are inevitably unstable~\cite{Khoury:2008wj}. The contracting phase in the original ekpyrotic scenario~\cite{Khoury:2001wf}, with $V(\phi) = -V_0e^{-c\phi/M_{\rm Pl}}$, is an attractor~\cite{Gratton:2003pe,Creminelli:2004jg},  as mentioned above, and correspondingly the growing mode for $\zeta$ is a constant in the long wavelength limit. However, the resulting spectrum for $\zeta$ is strongly blue~\cite{Lyth:2001pf,Brandenberger:2001bs,Gratton:2003pe,Creminelli:2004jg}, rather than scale invariant. A scale invariant spectrum can be obtained through entropy perturbations~\cite{Notari:2002yc,Finelli:2002we,Lehners:2007ac,Buchbinder:2007ad,Creminelli:2007aq}, as in the New Ekpyrotic scenario~\cite{Buchbinder:2007ad,Buchbinder:2007tw,Buchbinder:2007at}, but this requires two scalar fields. Even in this case, the entropy direction is generically tachyonically unstable~\cite{Koyama:2007mg,Tolley:2007nq}.

\begin{figure}
   \centering
   \includegraphics[width=4in]{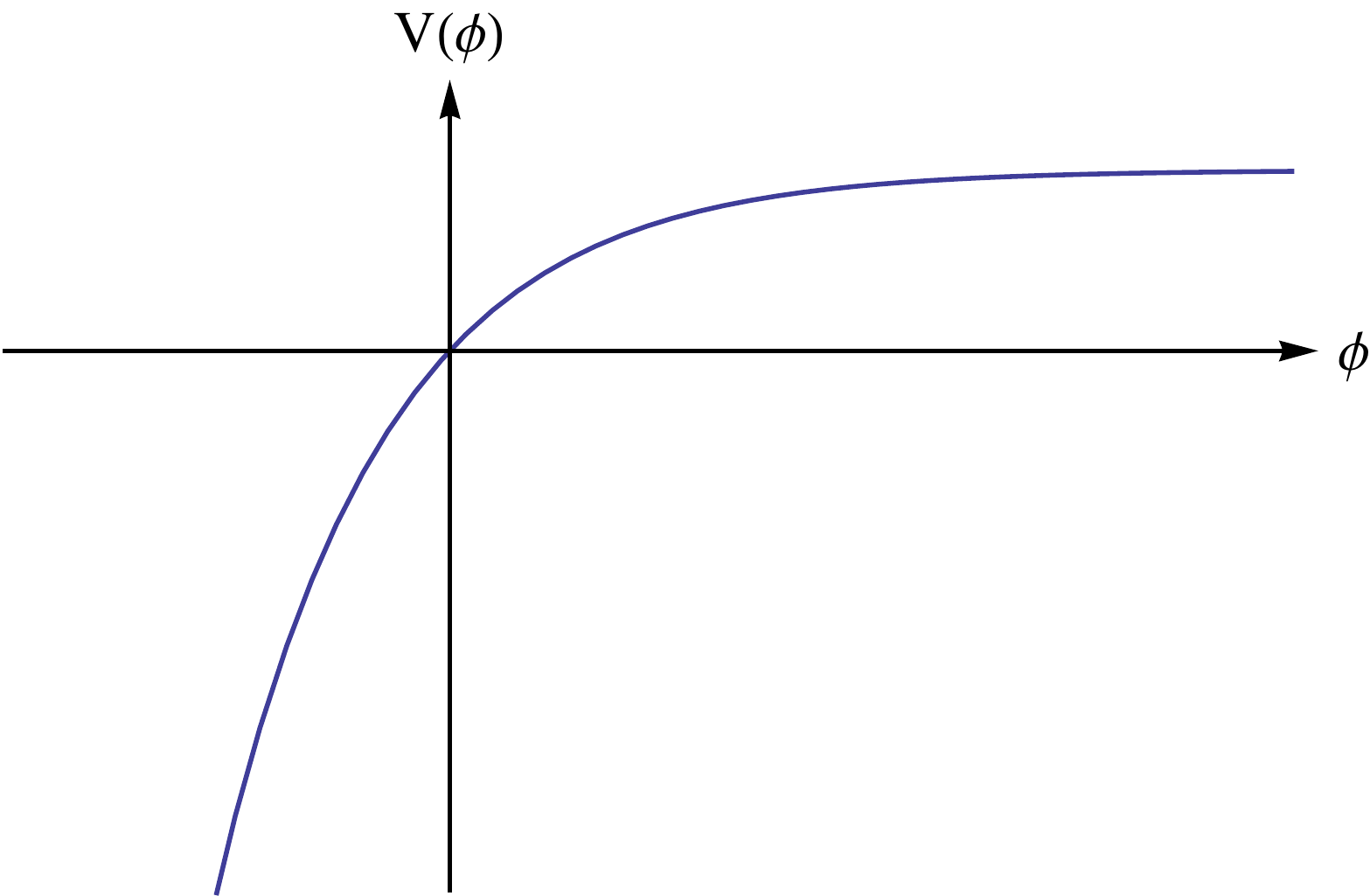}
   \caption{Depiction of the ``lifted exponential" potential, $V(\phi) = V_0 (1-e^{-c\phi/M_{\rm Pl}})$. At large field values the potential is nearly constant, and there is a steep waterfall around $\phi =0$.}
   \label{potentialcartoon}
\end{figure}

The recently-proposed {\it adiabatic ekpyrotic} mechanism~\cite{Khoury:2009my,Linde:2009mc,Khoury:2011ii,Kinney:2010qa}, based on a single scalar field with canonical kinetic term,
offers a counterexample. The mechanism can be realized with fairly simple potentials, such as the ``lifted exponential"
\be
V(\phi) = V_0 (1-e^{-c\phi/M_{\rm Pl}})\,,
\label{scalarpotential}
\ee
with $V_0 > 0$ and $c\gg 1$. This potential takes the form of a plateau that has been lifted to positive energy at large positive $\phi$ with a steep waterfall around $\phi=0$. See Fig.~\ref{potentialcartoon}. The key difference compared to earlier renditions of ekpyrotic cosmology is that the potential assumes {\it positive} values for part of the evolution.

Scale invariant adiabatic perturbations are generated during a transient phase in which the equation of state parameter,
\be
\epsilon \equiv -\frac{\dot{H}}{H^2} = \frac{3}{2}(1+w)\,,
\ee
grows rapidly from $\epsilon\ll 1$, where the constant term dominates, to $\epsilon \approx c^2/2 \gg 1$, where the negative exponential term dominates. 
During this transition, the scale factor is nearly constant, while the equation of state parameter varies rapidly as $\epsilon\sim 1/\tau^2$, where $\tau$ is conformal time.
This is referred to as the {\it transition phase}. The quantity $z \equiv a(\tau)  \sqrt{2 \epsilon(\tau)}$, which determines the evolution of $\zeta$, therefore scales as $z  \sim (-\tau)^{-1}$ --- exactly as in inflation, where $\epsilon\approx {\rm const.}$ and $a(\tau) \approx 1/(-\tau)$! The two-point function is, therefore, identical to inflation. (Another counterexample proposed
recently relies on a rapidly-varying, superluminal sound speed $c_s(\tau)$~\cite{csm,csk,csfedo}. See~\cite{Khoury:2008wj,csearlier,Piao:2006ja,Piao:2008ip} for earlier related work.) 

More generally, it has recently been shown that in fact there are only three single-field cosmologies with unit sound speed capable of generating a scale invariant spectrum for $\zeta$ on an attractor background~\cite{Khoury:2010gw,Baumann:2011dt}:  $i)$ inflation, with $a(\tau)\sim 1/|\tau|$ and $\epsilon\approx {\rm constant}$; $ii)$ the adiabatic ekpyrotic mechanism~\cite{Khoury:2009my,Khoury:2011ii} mentioned above,
with $\epsilon \sim 1/\tau^2$ on a slowly contracting background; and $iii)$ a novel version of the adiabatic ekpyrotic mechanism, in which the background first slowly expands, then slowly contracts. 
The analysis has been generalized to the case of a time-dependent sound speed~\cite{Baumann:2011dt,austin}. See~\cite{Piao:2003ty,Piao:2004uq,Kinney:2010qa} for related work.

Although all three cosmologies yield identical two-point function, the degeneracy is broken by the three-point function. In contrast with the extremely Gaussian spectrum
of the simplest inflationary models, the rapid variation of $\epsilon$ in the adiabatic ekpyrotic mechanisms leads to strongly scale-dependent non-Gaussianities, which
peak on small scales~\cite{Khoury:2011ii}. For the lifted exponential potential~(\ref{scalarpotential}), this results in a breakdown of perturbation theory on small scales, both classically and quantum mechanically~\cite{Khoury:2011ii}. As shown in~\cite{Khoury:2011ii}, however, these pathologies, and related issues raised in~\cite{Linde:2009mc}, can be avoided by considering more general forms of the potential, such as 
\be
V\left(\phi\right)=V_{0}\left(1-e^{-c(\phi)\phi/M_{\rm Pl}}\right)\,, 
\label{genpot}
\ee
where the exponent decreases from $c$ to a much smaller value $b\ll c$ after an acceptable range of scale invariant modes has been generated. 
The resulting spectrum is perturbative on all scales, but, because of the modified evolution, is now scale invariant over a finite range of modes, spanning a factor of $10^5$ in $k$ space, or a dozen e-folds.
While limited, this range is sufficient to account for observations of the cosmic microwave background and the large scale structure.

Earlier analyses of the adiabatic ekpyrotic mechanism have focused on the case where the universe is contracting throughout. In this paper, we instead explore the 
phenomenology and implications of the expanding case, consisting of an ``expanding transition phase", followed by a contracting ekpyrotic scaling phase. In doing so, we are motivated by the issue of embedding this mechanism in broader cosmological scenarios. Specifically, one attractive feature of the expanding branch is that we expect its basin of attraction to be much broader
than in the original contracting version. Indeed,  while the contracting transition phase in the original adiabatic ekpyrotic mechanism is a dynamical attractor, the evolution prior to the transition phase is generally not. At sufficiently early times, the field lies on a flat plateau of its potential, and, because the universe is contracting, any additional amount of kinetic energy is blueshifted and threatens to dominate the energy density. In the initially expanding version studied here, any additional kinetic energy will instead redshift away, and we therefore expect the solution to be stable for all times.

After reviewing the contracting version of the adiabatic ekpyrotic mechanism in Sec.~\ref{contractadiabatic}, we show in Sec.~\ref{expanding} that this mechanism works
equally well on a slowly expanding background and generates a scale invariant spectrum of curvature perturbations. Since scale invariance relies on a rapidly-varying
equation of state parameter while the scale factor is nearly constant, the density perturbation spectrum is to a good approximation insensitive to whether the background
is expanding or contracting during mode production. We check this explicitly in Sec.~\ref{numver} by numerically integrating the mode function equation in both the expanding and contracting
cases. The resulting spectra are indistinguishable (see Fig.~\ref{analyticzeta} for a preview). Section~\ref{rest} revisits the analytic calculation of the background evolution, allowing for
more general initial conditions. We quantify the extent to which the scale invariant phase shortens as a function of initial scalar field kinetic energy. In Sec.~\ref{phasespace},
we study the issue of stability globally by performing a phase space analysis, including a wide range of initial conditions, both for the expanding and the contracting branch.
This confirms that the transition phase is an attractor in both cases, as indicated at the perturbative level by $\zeta$ having a constant growing mode as $k\rightarrow 0$. More broadly, this analysis
also shows that the basin of attraction is broader in the expanding case, as anticipated in the previous paragraph. In Sec.~\ref{fusion} we study the connection to
an inflationary precursor to the transition phase, which follows in the expanding case from trusting~(\ref{scalarpotential}) at sufficiently large field values.
We briefly summarize our main results and discuss prospects for future directions in Sec.~\ref{conclu}.

The expanding adiabatic ekpyrotic mechanism studied here suffers from the same strong coupling issues as the original contracting version. In particular, the calculation
of the three-point amplitude in~\cite{Khoury:2011ii} neglected the time-dependence of the scale factor, and hence should apply to the expanding case as well. (One technical difference
in our case is that $\epsilon$ momentarily blows up at the end of the transition phase, because $H=0$ at that time. As the numerical analysis of Sec.~\ref{numver} clearly shows, however,
the two-point function is insensitive to this momentary divergence. We expect that the same is true for the three-point function.) In particular, just like in the contracting branch,
strong coupling can be avoided by considering more general forms of the potential, such as~(\ref{genpot}). We will not repeat the discussion of non-Gaussianities in this paper, and we refer
the interested reader to~\cite{Khoury:2011ii} for more details.

\section{Review of the Adiabatic Ekpyrotic solution}%%%%%%%%%%%%%%%%%%%%%%%%%%%%%%%%%%%%%%%%%%%%%%%%%
\label{contractadiabatic}

It is instructive to review the mechanism presented by~\cite{Khoury:2009my,Khoury:2011ii}~in which scale-invariant curvature perturbations
are generated by a fast-rolling scalar field during a slowly contracting phase. Consider a canonical scalar field
coupled to gravity
\be
S=\int d^4 x\sqrt{-g}\left(\frac{M_{\rm Pl}^2 R}{2}-\frac{1}{2}g^{\mu\nu}\partial_\mu\phi\partial_\nu\phi-V(\phi)\right)~,
\label{scalaraction}
\ee
where $M_{\rm Pl}^{-2} = 8\pi G$. For concreteness, we focus on the ``lifted exponential" potential~(\ref{scalarpotential}). 
This form of the potential is simplest for analytical calculations but is not meant to represent a realistic scenario. Indeed, to avoid strong coupling issues~\cite{Linde:2009mc}, realistic models must consider a more general potential of the form $V\left(\phi\right)=V_{0}\left(1-e^{-c(\phi)\phi/M_{\rm Pl}}\right)$, where $c(\phi)$ decreases to a smaller value after a suitable range of scale invariant modes has been generated. See~\cite{Khoury:2011ii} for details. Since the production of these scale-invariant modes occurs as the field traverses a small range $\Delta\phi\ll M_{\rm Pl}$, there is
a reasonable amount of freedom in specifying the global form of the potential.

The evolution of the scalar field on a cosmological background is governed by the equation of motion
\be
\ddot\phi+3H\dot\phi+V_{,\phi}=0\ .
\label{scalareom}
\ee
During the phase of interest, the background evolves extremely slowly so we may ignore Hubble friction. As shown in~\cite{Khoury:2009my}, this
approximation can be rigorously justified {\it a posteriori} by explicitly computing the first order correction and verifying that it is indeed small. 
Intuitively, the field falls down a steep waterfall, hence the transition occurs within a Hubble time. 
This ``fast roll" approximation leads to the simplified equation
\be
\ddot\phi\approx-V_{,\phi}=-\frac{c}{M_{\rm Pl}}V_{0}e^{-c\phi}\ ,
\label{scalareomapprox}
\ee
which is solved by 
\be
\phi(t)\approx\frac{2M_{\rm Pl}}{c}\log\left(\sqrt{\frac{V_{0}}{2M_{\rm Pl}^2}}c|t|\right)\ ~,
\label{phisol}
\ee
where $-\infty<t<0$. In writing down this solution, we have assumed that the field initially has negligible kinetic energy at early times, hence the total energy is $V_0$.
In Sec.~\ref{rest}, we will consider departures from this choice and the impact on the spectrum of perturbations.

To solve for the evolution of the background, we substitute~(\ref{phisol}) into $\dot H=-\dot\phi^2/2M_{\rm Pl}^2$ to obtain
\be
H(t)=\frac{2}{c^2t}-H_{0}\ .
\ee
At sufficiently early times, the Hubble parameter is approximately constant, $H(t)\simeq H_0$, where the
integration constant $H_0$ is fixed by the Friedmann equation,
\be
3M_{\rm Pl}^2H^2_0 \simeq  \frac{1}{2}\dot\phi^2 + V(\phi) \approx V_0\,.
\label{fried}
\ee
The authors of~\cite{Khoury:2009my} focused on the contracting branch,
\be
H_0=\sqrt{\frac{V_0}{3M_{\rm Pl}^2}}\ .
\ee
Indeed, since $H_0 > 0$ and $t$ is negative, $H(t)$ is manifestly negative, corresponding to a contracting universe.
(Note that the present sign convention differs from~\cite{Khoury:2009my}, where
$H(t) = 2/c^2t + H_0$ and $H_0$ is a negative quantity.)

The phase of interest is the regime where the $H_0$ term dominates the Hubble parameter, $H(t)\sim H_0$, in which case $\epsilon = -\dot{H}/H^2 \sim 1/t^2$.
Additionally, the assumption that the background remains nearly static implies $a(t) \sim 1$. The quantity $z \equiv a(\tau)  \sqrt{2 \epsilon(\tau)}$, which determines the
evolution of $\zeta$, therefore satisfies
\be
z\equiv a\sqrt{2\epsilon}\sim\frac{1}{\tau}~,
\ee
exactly as in inflation, where $\epsilon\approx {\rm const.}$ and $a(\tau) \approx 1/(-\tau)$.
Moreover, as in inflation, the growing mode for $\zeta$ goes to a constant, indicating that the background is a dynamical attractor~\cite{weinbergzeta}.
The two cosmologies yield identical power spectra, and therefore can be considered
``dual" to one another at the level of the two-point function. 
%With this basic understanding of the contracting case, we consider extending this mechanism to an expanding background.

The transition phase ends when $H(t)\simeq {\rm const.}$ is no longer a good approximation. 
Subsequently, the Hubble parameter tends to an ekpyrotic scaling
\be
H_{\rm ek}(t) \approx \frac{2}{c^2 t}~,
\label{Hek}
\ee
while the scale factor slowly decreases as
\be
a_{\rm ek}(t) \sim (-t)^{2/c^2}~.
\label{aek}
\ee
This ekpyrotic scaling phase is a necessary component of the story. Because the Hubble horizon is nearly static during the transition phase, the scale invariant modes
created in this phase remain inside the Hubble radius. A subsequent phase is therefore necessary to push these modes outside the Hubble horizon.
The scaling ekpyrotic phase fills this role --- since the universe is slowly contracting, modes are gently pushed outside the Hubble horizon without disturbing their spectrum.
(Eventually, the ekpyrotic phase must itself terminate before the big crunch. We envision that it is followed by a bounce to an expanding, radiation-dominated phase. At the level of 
a four-dimensional effective theory, a stable non-singular bounce can be achieved either through a phase of ghost condensation~\cite{Creminelli:2006xe}, as in the New Ekpyrotic scenario,
or through a phase of galileon domination~\cite{Nicolis:2009qm}. See~\cite{Khoury:2010gb,Khoury:2011da} for recent supersymmetric extensions of these theories.)

\section{Adiabatic Mechanism in an Expanding Phase}
\label{expanding}

The key element of the adiabatic ekpyrotic mechanism is the rapid evolution of the equation of state parameter $\epsilon$; this growth is responsible for the generation of a spectrum of adiabatic modes. 
Interestingly, the scale factor remains nearly constant in the process, and hence plays no essential role. As a result, we expect that the adiabatic ekpyrotic mechanism can be generalized
to a case where the scale factor is slowly \textit{increasing}. In this Section we show that this is indeed possible.

Though this may seem like an academic exercise, there are important implications in extending the original set-up,
especially for embedding this mechanism in broader cosmological scenarios. As reviewed above, in the contracting
adiabatic mechanism the transition phase is a dynamical attractor~\cite{Khoury:2011ii}. However, the cosmological
evolution prior to the onset of the transition phase is not necessarily an attractor solution. In fact, we know that
at the earliest times when the field lies on the potential plateau, the background cosmology must be unstable ---
any additional kinetic energy in the field will blueshift and can lead to kinetic domination. In an initially expanding universe, 
the kinetic energy will instead redshift. In this way, we expect an expanding transition phase solution to be stable for all time.

\subsection{The Expanding Transition Solution}

The construction of the solution is quite similar to the contracting case. As before, we may ignore the background evolution to first order so the scalar field equation of motion reduces to~(\ref{scalareomapprox}), with a solution given again by
\be
\phi(t)\approx\frac{2M_{\rm Pl}}{c}\log\left(\sqrt{\frac{V_{0}}{2M_{\rm Pl}^2}}c|t|\right)\ ~,
\label{phisol2}
\ee
where $-\infty<t<0$. Since the cosmological background is to first order irrelevant for the scalar field evolution, it is natural that the approximate solution~(\ref{phisol}) applies
irrespective of whether the universe is contracting or expanding.  In the same way as before, we can integrate the $\dot H$ equation to obtain
\be
H(t)=\frac{2}{c^2t}+H_0\ .
\label{H(t)}
\ee
The Friedmann equation~(\ref{fried}) constrains the magnitude of the integration constant $H_0$ as before, 
\be
H_0=\sqrt{\frac{V_0}{3M_{\rm Pl}^2}}\ .
\label{H0}
\ee
Note this time we have chosen the positive-definite quantity $H_0$ to appear as $+H_0$ in~(\ref{H(t)}). Hence, at sufficiently early times, 
$H(t)\approx H_0$ is approximately constant and positive, corresponding to an expanding de Sitter universe. 
The expanding transition solution will be similar to the contracting case --- as long as $H(t)\approx H_0$, we will have $z\sim 1/\tau^2$, and
scale invariant perturbations will be generated. However since Hubble is positive during this time, perturbations are
generated on a slowly \textit{expanding} background. 

The transition phase is once again followed by an ekpyrotic scaling phase, with $H_{\rm ek}(t) \approx 2/c^2 t$.
During this phase, the Hubble radius shrinks slowly, and the scale invariant modes are gently pushed
outside the horizon. There is, however, a key difference: in the expanding case, $H(t)$ changes sign as the universe evolves from the (expanding) transition phase
to the (contracting) ekpyrotic scaling phase. 
%We will check in Sec.~\ref{numver} that the momentary blow-up of the Hubble radius when $H=0$
%does not affect the curvature perturbation. The spectrum of perturbations in the two branches are in fact indistinguishable.

By definition, the transition phase proceeds as long as the Hubble parameter is nearly constant and the Hubble friction term can be
neglected in the scalar field evolution. These conditions are only satisfied for a finite time. First note that the constant term dominates in the expression for $H(t)$ until
\be
t_{\rm end-tran}=t_{\rm beg-ek} =-\frac{2}{c^2H_0}\ ,
\label{tend}
\ee
which corresponds to the time when $H$ vanishes, and the universe transitions from expansion to contraction.
As the subscripts indicate, this marks the end of the transition phase and the onset of the ekpyrotic scaling phase.
Likewise, the transition phase is also finite in the past. The solution~(\ref{phisol2}) for $\phi(t)$ neglected gravity,
which is a poor approximation for sufficiently large positive $\phi$ where the potential is flat and Hubble damping
is important. More precisely, it is straightforward to show that the approximation $H\dot{\phi} \ll c V_0e^{-c\phi/M_{\rm Pl}}/M_{\rm Pl}$ used
in~(\ref{scalareomapprox}) is consistent as long as $t > t_{\rm beg-tran}$, where
\be
t_{\rm beg-tran} =-\frac{1}{H_{0}}\ .
\label{tbeg}
\ee
Summarizing, the transition phase solution is valid for
\be
-\frac{1}{H_0}=  t_{\rm beg-tran} < t < t_{\rm end-tran} = -\frac{2}{c^2H_0}~.
\ee
Interestingly, the length of the transition phase lasts less than a Hubble time, and thus represents a small amount of the total cosmological evolution.

We can calculate the equation of state parameter $\epsilon$ using our expression for $H$ in~(\ref{H(t)}):
\be
\epsilon= - \frac{\dot{H}}{H^2} =\frac{c^2/2}{\left(1+H_{0}c^2 t/2\right)^2}=\frac{c^2/2}{\left(1-t/t_{\rm end-tran}\right)^2}\ .
\label{epsexpand}
\ee
At early times, $|t| \gg |t_{\rm end-tran}|$, this expression reduces to $\epsilon\sim1/t^2$, characteristic of the transition phase.
Hence we expect that this will lead to the generation of scale invariant modes. At late times, $|t| \ll |t_{\rm end-tran}|$, we have
$\epsilon\approx c^2/2$, characteristic of the ekpyrotic scaling phase. 

Note that the equation of state parameter diverges at $t = t_{\rm end-tran}$ --- the Hubble parameter vanishes at that time while $\dot H$ remains finite. 
At first sight this may seem worrisome, since the momentary divergence of $\epsilon$ could potentially disrupt the scale invariance of the perturbations. 
This turns out to be an unfounded fear. We will show in Sec.~\ref{numver} by numerically integrating the perturbation equation that the spectrum
for $\zeta$ is unaffected by the momentary blow-up in the equation of state. Intuitively, this is because the time scale over which the equation of
state diverges is very short, {\it i.e.} much less than a Hubble time.

For completeness, we can integrate~(\ref{H(t)}) to obtain an expression for the scale factor
\be
a(t)\simeq (-t)^{2/c^2}e^{H_0t}\simeq 1+\frac{2}{c^2}\log(-t)+H_0t+\ldots\ ,
\label{aexpand}
\ee
where the Taylor expansion is a good approximation during the transition phase. The scale factor is therefore nearly constant during the transition
phase, as in the contracting example, and remains finite throughout the evolution. 
  
It is worth emphasizing that while the background is expanding during the transition phase, this is decidedly not an inflationary scenario. During the expanding transition phase,
the equation of state evolves rapidly in time ($\epsilon(t) \sim1/t^2$), while the scale factor is nearly constant $(a(t)\approx 1)$, in stark contrast to slow-roll inflation.\\

\subsection{Curvature Perturbations in the Expanding Phase}
\label{exppert}

We are now in a position to check that a slowly-expanding transition phase leads to a scale invariant spectrum of density perturbations. We begin with an analytic calculation,
before verifying the results numerically in Sec.~\ref{numver}. For this purpose, we work in comoving gauge, where the spatial slices are constant density ($\delta\phi=0$)
hypersurfaces, and the spatial metric is given by $h_{ij}=a^2(t)e^{2\zeta}\delta_{ij}$. In this gauge, the variable $\zeta$ represents the curvature perturbation on spatial slices.
Its action at quadratic order is given by
\be
S_{2}=\frac{1}{2}\int d^3xd\tau~z^2\left[\zeta'^2-\left(\vec\nabla\zeta\right)^2\right]~,
\ee
where $z = a\sqrt{2\epsilon}$ as before, and primes denote derivatives with respect to conformal time $\tau$. The resulting equation of motion for the Fourier modes $\zeta_k$ is
\be
\zeta_k''+2\frac{z'}{z}\zeta_k + k^2\zeta_k=0\,.
\label{zetaeom}
\ee
It is convenient to instead work in terms of the canonically normalized variable $v\equiv z\cdot\zeta$:
\be
v_{k}''+\left(k^2-\frac{z''}{z}\right)v_{k}=0\ .
\label{modefunction}
\ee
If $z''/z\sim 2/\tau^2$, as in inflation, then this equation is well-known to yield a scale-invariant spectrum for $\zeta$ at long wavelengths. Furthermore, in this case the growing mode of $\zeta$ goes to a constant as $k\rightarrow 0$. In this limit, $\zeta$ may be interpreted as a homogeneous perturbation to the scale factor and may be locally absorbed by a spatial diffeomorphism \cite{weinbergzeta}.
The background is therefore a dynamical attractor.

In our case, it follows from~(\ref{epsexpand}) and~(\ref{aexpand}) that
\be
z(t)=(-t)^{2/c^2}e^{H_{0}t}\frac{c}{1-t/t_{\rm end-tran}}\ .
\label{zee}
\ee
Thus $z$ inherits the singular behavior of $\epsilon$ at $t = t_{\rm end-tran}$, though as mentioned earlier this will not pose a problem.
One way to see this is to note that $t=t_{\rm end-tran}$ is a regular singular point of~(\ref{zetaeom}). 

During the transition phase, the scale factor is nearly constant, hence conformal time and cosmological time are approximately
the same: $t\approx \tau$. Deep in the transition phase, $|t| \gg |t_{\rm end-tran}|$,~(\ref{zee}) therefore implies
\be
\frac{z''}{z}\approx\frac{2}{(t-t_{\rm end-tran})^2}\approx\frac{2}{\tau^2}\,.
\ee
Assuming the usual adiabatic vacuum, the solution is
\be
v_{k}=\frac{e^{-ik\tau}}{\sqrt{2k}}\left(1-\frac{i}{k\tau}\right)\,,
\ee
which gives a scale invariant power spectrum for the curvature perturbation, $\zeta= v/z$, in the long wavelength limit ($k|\tau|\ll 1$):
\be
P_{\zeta}\equiv\frac{1}{2\pi^2}k^3|\zeta_{k}|^2=\frac{c^2V_0}{48\pi^2}\ .
\label{power}
\ee
Moreover, in this limit $\zeta$ approaches a constant, indicating that the expanding transition phase is also an attractor. 

Since modes freeze out when $k|\tau|\approx k|t|\sim1$, the range of scale invariant modes is set by the duration of the transition phase.
From~(\ref{tend}) and~(\ref{tbeg}), we deduce that
\be
\frac{k_{\rm max}}{k_{\rm min}}\approx\frac{t_{\rm beg-tran}}{t_{\rm end-tran}}=\frac{c^2}{2}\ .
\label{length}
\ee
Equations~(\ref{power}) and~(\ref{length}) agree exactly with the contracting case~\cite{Khoury:2009my,Khoury:2011ii}. Of course, this is not surprising, since away from the singular point $t = t_{\rm end-tran}$, $z$ does not know the difference between expanding and contracting solutions. We now check such claims through numerical integration.

\section{Numerical Verification of Scale Invariance}
\label{numver}

In this Section we check the analytical results derived above by numerically integrating the evolution of $\zeta$.
In the process we will reassure ourselves that the momentary singularity in $\epsilon$ does not spoil the scale invariance of the
modes created during the transition phase. 

Our starting point is the evolution equation~(\ref{zetaeom}) for the mode functions $\zeta_k$. This equation is better behaved than that for $v_k$,
since, as mentioned earlier, the singularity at $t = t_{\rm end-tran}$ is a regular singular point in this case. As initial conditions we impose the
adiabatic vacuum choice 
\be
\zeta_k(\tau_i)=\frac{1}{z(\tau_i)\sqrt{2k}}e^{ik\tau_i}~,
\ee
with $\tau_i$ chosen so that $k|\tau_i| \gg 1$ for all modes of interest.

To proceed, we need to substitute an expression for $z(\tau)$ into~(\ref{zetaeom}). We will do this in two ways. First, following a quasi-analytic approach,
in Sec.~\ref{quasianalytic} we use the analytic form given by~(\ref{zee}). This expression is of course not exact, since various approximations went into deriving it.
Second, in Sec.~\ref{exact} we will redo the calculation more precisely using an exact form for $z(\tau)$, which will be obtained by numerically integrating the
background equations of motion.

\subsection{Integration with Analytic $z(\tau)$}
\label{quasianalytic} 

Let us start with the quasi-analytic approach, using the approximate form for $z(\tau)$ given by~(\ref{zee}). As it stands,~(\ref{zee}) gives
$z$ as a function of cosmological time $t$, whereas we need it in terms of conformal time $\tau$. Note that the factors of $c/(1-t/t_{\rm end-tran})$ and $(-t)^{2/c^2}$
are important in the transition phase and the ensuing contracting ekpyrotic phase, respectively. During both of these phases, cosmological time and conformal time are nearly the same, so a good approximation is to replace $t$ with $\tau$ in these factors. The factor of $e^{H_{0}t}$, however, is important at early times and makes the scale factor differ from unity --- conformal time and proper time 
are therefore much different in this case. Approximating $a(t) \simeq e^{H_0 t}$ at early times, we can integrate $d\tau=dt/a(t)$ to obtain
\be
e^{H_{0}t}=\frac{H_{0}^{-1}}{H_{0}^{-1}-\tau}=\frac{1}{1-H_{0}\tau}\ .
\ee
Thus we obtain the following analytic expression for $z(\tau)$
\be
z(\tau)=\frac{c(-\tau)^{2/c^2}}{\left(1-H_{0}\tau\right)\left(1-\tau/t_{\rm end-tran}\right)}\ .
\label{ztau}
\ee

Using this analytic expression for $z(\tau)$ we are now in a position to integrate the equation of motion for $\zeta$. In fact, we will do the integration for both the expanding and the contracting
case, where the contracting case corresponds to letting $H_0 \rightarrow -H_0$ in~(\ref{ztau}). In both cases, we take $c=200$, $H_0=5\times 10^{-4}$, and integrate~(\ref{zetaeom}) over the range of modes $.02H_{0}<k<2\times 10^4H_{0}$ and over the time interval $-.5H_{0}^{-1}<\tau<-5\times 10^{-10}H_{0}^{-1}$. These parameters are chosen so that the shortest wavelength
modes will have just left the Hubble horizon by the end of the evolution, while the background solution will be firmly within the scaling ekpyrotic phase.

\begin{figure}
   \centering
   \subfigure[Expanding]{\includegraphics[width=3.4in]{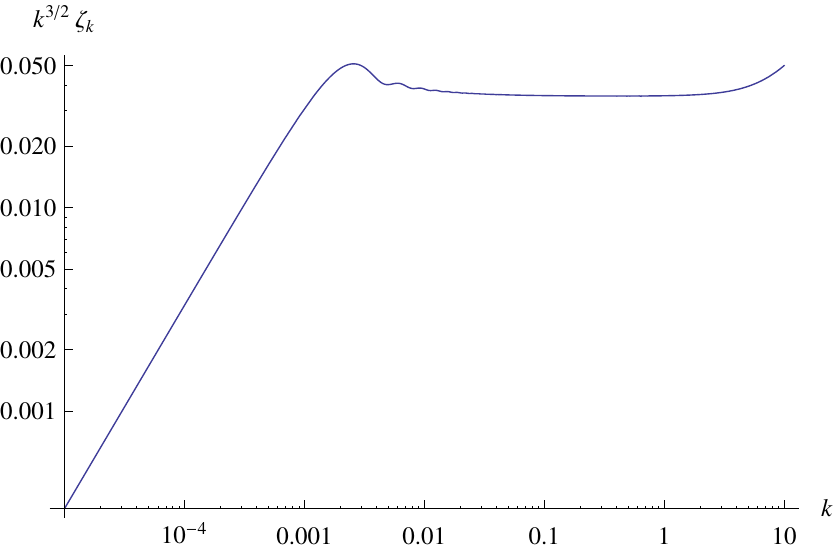}}
   \subfigure[Contracting]{\includegraphics[width=3.4in]{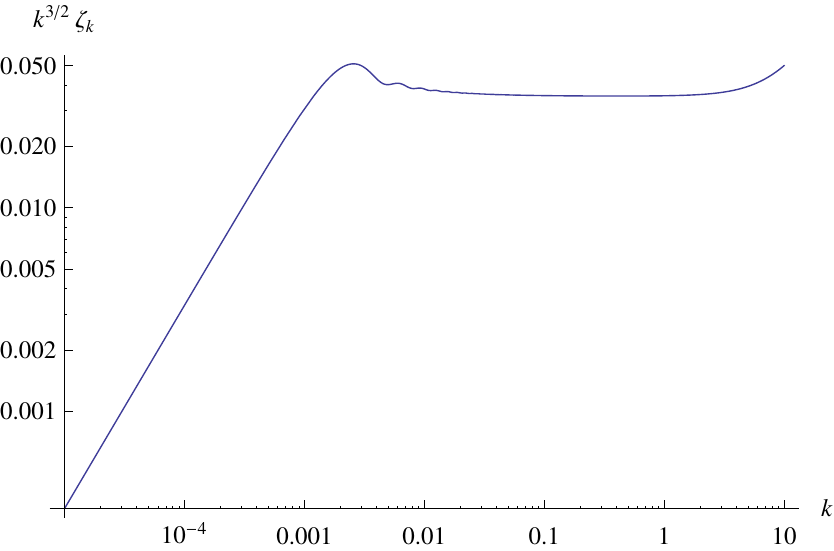}}
   \caption{Numerical computation of the power spectrum $k^{3/2}\zeta_k$ vs. $k$ in both the expanding and contracting transition phase scenarios using an analytic expression for $z(\tau)$.  This confirms that the spectrum is insensitive to whether the transition phase is contracting or expanding.}
   \label{analyticzeta}
\end{figure}

The resulting power spectra are shown in Fig.~\ref{analyticzeta}. We find that there is a range of scale-invariant modes spanning roughly $4$ decades in $k$-space, which is in order-of-magnitude agreement with our analytical result (\ref{length}). Note that the range of scale invariant modes is slightly shorter than that found in the numerical analysis of~\cite{Khoury:2009my}, which is due to our including the extra factor $1/(1-H_0\tau)$ in $z(\tau)$. We see that the power spectrum for $\zeta$ is indistinguishable in each case, which confirms that the adiabatic ekpyrotic
generation mechanism works equally well on an expanding background. In particular, this confirms that the divergence in $\epsilon$ at $t=t_{\rm end-tran}$, corresponding to the transition from expansion to contraction, has no effect on the perturbation spectrum.

\subsection{Numerical Solution of the Full Background Equations}
\label{exact}

The above analysis relied on certain approximations in deriving (\ref{ztau}). We neglected Hubble friction in the evolution for $\phi$, but this must eventually break down at
early times when the background is approximately expanding de Sitter space. To check that such corrections to~(\ref{ztau}) do not spoil scale invariance, in this Section 
we numerically solve for the background evolution to obtain an exact result for $z(\tau)$.

This can be done using the Hamilton-Jacobi formalism. Using the chain rule, the $\dot{H}$ equation implies
\be
H_{,\phi}=-\frac{\dot\phi}{2M_{\rm Pl}^2}\,,
\label{Hphi}
\ee
hence the Friedmann equation can be rewritten as
\be
3M_{\rm Pl}^2H^2=\frac{1}{2}\dot\phi^2+V(\phi)=2M_{\rm Pl}^2 H_{,\phi}^2+V(\phi)\ .
\label{HJ}
\ee
In this way, $\phi$ is thought of as the clock tracking the background evolution. We first numerically solve this equation to find $H(\phi)$,
and in turn obtain $\phi(t)$ by integrating~(\ref{Hphi}). At the end of the day, this gives us the Hubble parameter as a function of time $H(\phi(t))$,
from which we can extract $a(t)$. In practice, we perform the integration over a sufficiently broad range of field values and correspondingly
large time interval. Specifically, we have solved~(\ref{HJ}) over the field range $0.07\;M_{\rm Pl} \leq \phi \leq -M_{\rm Pl}$, setting
$H(0.07\;M_{\rm Pl}) = H_0$. Meanwhile, in integrating~(\ref{Hphi}) to obtain $\phi(t)$ we fix the integration constant corresponding to a shift in time
by demanding that $\phi$ matches the analytic solution~(\ref{phisol2}).

\begin{figure}
   \centering
   \subfigure[Numerical $z(\tau)$]{\includegraphics[width=3.4in]{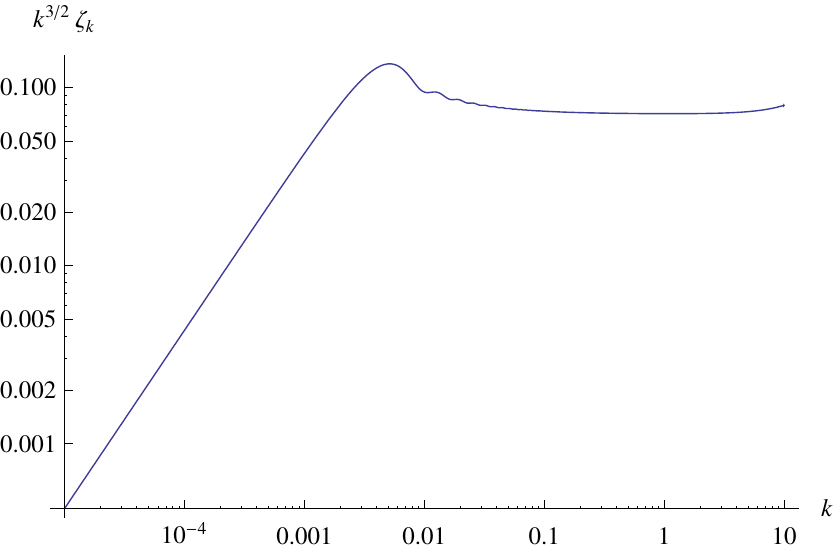}}
   \subfigure[Analytic $z(\tau)$]{\includegraphics[width=3.4in]{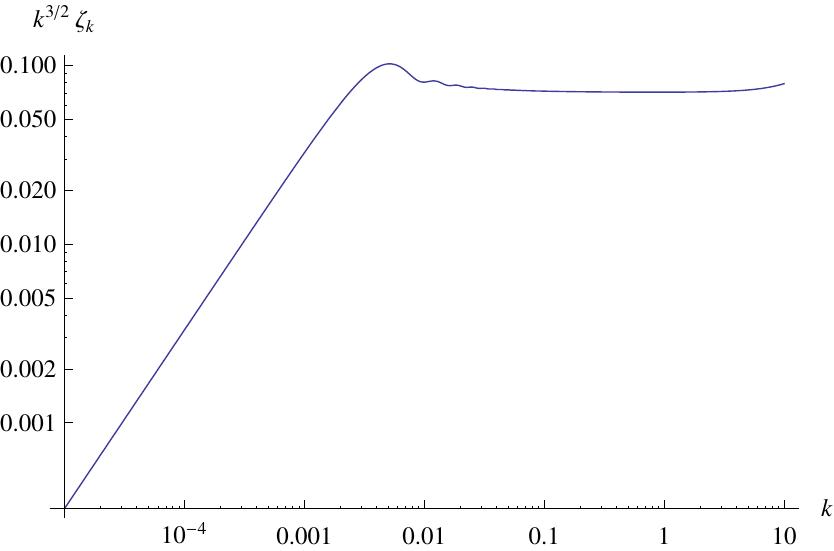}}
   \caption{Comparison of the power spectra for the exact numerical calculation and the integration using our analytic expression for $z(\tau)$.
   %Aside from the difference in normalization, 
   The curves are in excellent agreement.}
   \label{numericalzeta}
\end{figure}

Substituting everything into $z = a\sqrt{2\epsilon}$, we can numerically integrate~(\ref{zetaeom}) to obtain the power spectrum.
The result is shown in Fig.~\ref{numericalzeta}a. For comparison, Fig.~\ref{numericalzeta}b shows the quasi-analytic result.
Computational constraints forced us to use slightly different integration parameters from those used in Sec.~\ref{quasianalytic}.  
Figure \ref{numericalzeta} was obtained using $c =200$ and $H_0 = 10^{-3}$, with integration ranging over
$-.5H_0^{-1} < \tau < -10^{-8}H_0^{-1}$ and $.01 H_0 < k < 10^4 H_0$. 
We see that the plots show very good agreement.

\section{Starting From Rest}
\label{rest}

The analysis of Sec.~\ref{expanding} assumed a scalar field starting from rest ($\dot\phi=0$) in the asymptotic past. 
In this Section we explore a broader range of initial conditions, corresponding to non-zero initial kinetic energy for the
scalar field, and the impact on the range of scale invariant modes. Specifically, our goal is to derive an expression for the length of time spent
in the transition phase (and thus the range of scale invariant $k$-modes generated) as a function of the initial energy.

We start with the equation of motion for the scalar field, assuming as before that the Hubble damping term is negligible
\be
\ddot\phi+V_{,\phi}\approx 0\ .
\ee
This equation admits a first integral of motion, 
\be
\frac{1}{2}\dot\phi^2+V(\phi)=E\ ,
\label{integralofmotion}
\ee
where $E$ is of course the total energy of the field. In Sec.~\ref{expanding}, as well as in earlier work~\cite{Khoury:2009my,Khoury:2011ii}, $E$ was taken to be equal to $V_{0}$ at the onset of the transition phase, while remaining agnostic about the prior evolution.  Here we want to consider the more general case $E<V_{0}$. In other words, $E$ corresponds to the value of the potential from which the scalar field starts at rest. 

Equation~(\ref{integralofmotion}) implies
\be
\dot\phi=-\sqrt{2(E-V(\phi))}\ ,
\ee
where we have chosen the negative branch of the square root, corresponding to the field rolling downhill.
Substituting our potential~(\ref{scalarpotential}), this can be integrated explicitly
\be
-t= \int\frac{e^{c\phi/2M_{\rm Pl}}d\phi}{\sqrt{2\left[V_{0}+(E-V_{0})e^{c\phi/M_{\rm Pl}}\right]}} = \frac{2M_{\rm Pl}}{c\sqrt{2(V_0-E)}}\arcsin\left(\sqrt{\frac{V_0-E}{V_0}}e^{c\phi/2M_{\rm Pl}}\right) \,,
\ee
where the last step assumes $e^{c\phi/2} \leq \sqrt{V_0/(V_0-E)}$. This can be inverted 
to obtain a solution for $\phi(t)$, valid as long as $\phi\leq (2/c)\log{ \sqrt{V_0/(V_0-E)}}$:
\be
\phi(t)=\frac{2M_{\rm Pl}}{c}\log\left[\sqrt{\frac{V_0}{V_0-E}}\sin\left(\sqrt{\frac{V_0-E}{2M_{\rm Pl}^2}}c|t|\right)\right]\ .
\label{fullphi}
\ee
This expression generalizes~(\ref{phisol2}) to span a range of initial field energy, or equivalently initial position on the potential where the field is released from rest.
As a check, this reduces to~(\ref{phisol2}) in the limit $E \rightarrow V_0$. Moreover, at late times, when $|t|$ becomes sufficiently small,~(\ref{fullphi}) also
approaches~(\ref{phisol2}), confirming that this is an attractor.

It is straightforward to calculate $\dot H$
\be
\dot H=-\frac{1}{2M_{\rm Pl}^2}\dot\phi^2=-\frac{V_0-E}{M_{\rm Pl}^2}\cot^2\left(\sqrt{\frac{V_0-E}{2M_{\rm Pl}^2}}c|t|\right)\ .
\label{dotHgen}
\ee
Similarly, substituting the above solution for $\phi(t)$ in the Friedmann equation, $3H^2M_{\rm Pl}^2=\frac{1}{2}\dot\phi^2+V(\phi)=E$, with $V(\phi)=V_0(1-e^{-c\phi})$,
we obtain $H^2=E/3M_{\rm Pl}^2$. The equation of state parameter is therefore given by
\be
\epsilon=-\frac{\dot H}{H^2}=\frac{3(V_0-E)}{E}\cot^2\left(\sqrt{\frac{V_0-E}{2M_{\rm Pl}^2}}c|t|\right)\ .
\ee
This generalizes~(\ref{epsexpand}) to a broader range of initial conditions, and matches~(\ref{epsexpand}) in the limit $E \rightarrow V_0$, as well as at late times $|t|\rightarrow 0$.

Clearly the range of scale invariant modes will depend on the initial conditions, and we can already expect a
more restricted range as we move away from the case $E=V_0$ studied earlier. The relevant quantity to
assess the shape of the power spectrum is the time-dependent mass term, $z''/z$, appearing in the mode function equation~(\ref{modefunction}).
Since $a(t)$ is nearly constant during the phase of interest, we have
\be
\frac{z''}{z}\approx \frac{1}{\sqrt{\epsilon}} \frac{{\rm d}^2\sqrt{\epsilon}}{{\rm d}t^2} = \frac{c^2(V_0-E)}{M_{\rm Pl}^2}\csc^2\left(\sqrt{\frac{V_0-E}{2M_{\rm Pl}^2}}c|t|\right)\,.
\ee
Modes will be scale invariant provided they freeze out when $z''/z \approx 2/t^2$. Recalling the Taylor expansion $\csc^2x\simeq x^{-2}+1/3$, 
this will be the case whenever $c(-t)\sqrt{(V_0-E)/2}\ll 1$. This condition is approximately satisfied for $t < \widetilde t_{\rm beg-tran}$, where
\be
\widetilde t_{\rm beg-tran}\approx -\frac{1}{c}\sqrt\frac{2}{V_0-E}\ .
\label{beggen}
\ee
In other words, $\widetilde t_{\rm beg-tran}$ marks the onset of scale invariant mode production. Although this expression naively diverges for $E=V_0$, we of course can only make $\widetilde t_{\rm beg-tran}$ as large as $t_{\rm beg-tran} = -1/H_0$ --- at earlier times, the fast-roll approximation assumed here breaks down. More carefully, we have
\be
\widetilde t_{\rm beg-tran}\approx {\rm max}\left(-\frac{1}{c}\sqrt\frac{2}{V_0-E},\;-\sqrt{\frac{3M_{\rm Pl}^2}{V_0}}\right)~.
\ee
The scale invariant phase has finite duration. Much like the transition phase studied earlier, it comes to an end
when the approximation $H\simeq \sqrt{E/3M_{\rm Pl}^2}$ breaks down. Integrating~(\ref{dotHgen}) in the limit
$c(-t)\sqrt{(V_0-E)/2}\ll 1$ gives
\be
H(t)=\sqrt\frac{E}{3M_{\rm Pl}^2} + \frac{2}{c^2 t} +\ldots\ ,
\ee
where the ellipses include terms that become increasingly small as $t \rightarrow 0$.
It follows that the scale invariant or transition phase concludes at
\be
\widetilde t_{\rm end-tran}\approx -\frac{2}{c^2}\sqrt\frac{3M_{\rm Pl}^2}{E}\ .
\label{endgen}
\ee

Combining~(\ref{beggen}) and~(\ref{endgen}), the range of scale invariant modes is thus given by 
\be
\frac{k_{\rm max}}{k_{\rm min}}\approx\frac{\widetilde t_{\rm beg-tran}}{\widetilde t_{\rm end-tran}} < \frac{c^2}{2} \sqrt{\frac{E}{V_0}}\,.
\label{lengthnew}
\ee
Here we have assumed the earliest possible onset of the transition phase. Comparing to our previous answer~(\ref{length}), $k_{\rm max}/k_{\rm min} = c^2/2$, we see that
solutions with $E<V_0$ lead to a narrower range of scale invariant modes, as expected. Note that~(\ref{lengthnew})
agrees with~(\ref{length}) in the limit $E\rightarrow V_0$, as it should.

\section{Phase Space Analysis}
\label{phasespace}

In Sec.~\ref{exppert}, as well as in earlier work~\cite{Khoury:2009my,Khoury:2011ii}, the adiabatic ekpyrotic evolution has been argued to
be a dynamical attractor because the growing mode for $\zeta$ goes to a constant in the long wavelength limit~\cite{weinbergzeta}. This argument, while true, only applies
at the perturbative level, {\it i.e.} for sufficiently small deviations from the background solution. In this Section, we study the issue of stability
more broadly, by performing a phase space analysis for a wide range of initial conditions. This will allow us to determine the breadth of the basin
of attractor both in the expanding and in the contracting branches of the transition phase.

For this purpose, we consider curves in the $(\phi,\dot\phi)$ phase plane parameterized by $N\equiv\log a$. 
The Friedmann constraint and the scalar field equation of motion imply the autonomous system:
\begin{eqnarray}
\nonumber
\frac{{\rm d}\phi}{{\rm d}N}&=&\frac{\dot\phi}{H}\,; \\
\frac{{\rm d}\dot\phi}{{\rm d}N}&=&-3\dot\phi-\frac{V_{,\phi}}{H}~,
\label{phidotphivector}
\end{eqnarray}
where the Hubble parameter is understood as a function of $\phi$ and $\dot{\phi}$:
\be
H=\pm\frac{1}{M_{\rm Pl}}\sqrt{\frac{\dot\phi^2}{6}+\frac{V(\phi)}{3}}\,.
\ee
The choice of sign for the square root corresponds to the choice of expanding or contracting branch.
Cosmological solutions are given by the integral curves of the vector field
$({\rm d}\phi/{\rm d}N, {\rm d}\dot{\phi}/{\rm d}N)$. 

\subsection{Definition of Attractor}

Let us be precise about what we mean by attractor behavior. As curves approach each other in the $(\phi,\dot\phi)$ plane, their respective values of $\phi(N)$ and $\dot\phi(N)$ get closer and closer together, up to a constant relative shift in $N$.  We have the gauge freedom to fix the initial values of $N$ such that the solutions coincide. In this sense, an attractor solution is identified by neighboring curves flowing towards it.
We will see that this is the case for the transition phase and the subsequent ekpyrotic scaling phase. 

It is worth pointing out that this focusing of trajectories occurs in the $(\phi,\dot\phi)$ plane, but not in the physical phase space. 
Minisuperspace models are Hamiltonian systems, and consequently Liouville's theorem forbids reduction of phase space volume.
The resolution of this apparent discrepancy is of course that $p_\phi = a^3\dot\phi$, rather than $\dot\phi$, is the momentum conjugate to $\phi$
in the canonical formalism.

This can be seen more generally by considering a $2n$ dimensional phase space. Minisuperspace is a constrained Hamiltonian system, with the Friedmann equation defining a $2n-1$ dimensional constraint hypersurface on which the Hamiltonian vanishes. This hypersurface is foliated by the gauge orbits corresponding to the Hamiltonian flow. The $2n-2$ dimensional hypersurface which is transverse to this Hamiltonian flow is the space of classical trajectories; this is the symplectic reduction or Marsden-Weinstein quotient $M=\mathcal{H}^{-1}(0)/\mathbb{R}$, where $\mathcal H^{-1}(0)$ is the locus where the Hamiltonian vanishes~\cite{gibbonshawking,gibbonsturok}. There is a well-defined pullback of the symplectic form in the total space to this reduced phase space. Darboux's theorem tells us that there are local coordinates $(p_i, q^i)$ such that the symplectic 2-form is
\be
\omega = \sum_{i=1}^n {\rm d}p_i\wedge {\rm d}q^i~.
\ee
For a model with action (\ref{scalaraction}) in the ADM formalism we can choose coordinates to write this symplectic form as
\be
\omega =  {\rm d}\mathcal{H}\wedge {\rm d}t + \sum_{i=1}^{n-1} {\rm d}p_i\wedge {\rm d}q^i~.
\ee
In this language, the Friedmann equation is the constraint that the Hamiltonian vanishes $\mathcal{H}=0$. The total phase space carries a natural invariant measure constructed from $\omega$. We may pull back this symplectic form to the reduced phase space $M$ and construct a natural measure $\Omega$. This is the Gibbons Hawking Stewart measure \cite{gibbonshawking}.\\
\\
Note that on the reduced phase space, this measure is conserved. Moreover, in the particular case of a spatially-flat ($k=0$) universe we can write the equations of motion in such a way that they are independent of $a$, as we did in the single-field autonomous system (\ref{phidotphivector}). Focusing on this single-field case for concreteness, we may then choose to parameterize the reduced phase space by $\phi$ and $\dot\phi$ \cite{hollandswald}. Choosing to treat these as Euclidean coordinates introduces the measure
\be
\mu = {\rm d}\dot\phi\wedge {\rm d}\phi~,
\label{adhocmeasure}
\ee
but this of course need not be conserved by the evolution. As mentioned earlier, this non-conservation can be traced to the fact that $\dot\phi$ is not the momentum canonically conjugate to $\phi$.
Using~(\ref{adhocmeasure}) is nevertheless a sensible thing to do --- as mentioned above, we are interested in trajectories with the same values of $\phi$ and $\dot\phi$.

\begin{figure}
   \centering
   \subfigure[Expanding]{\includegraphics[width=3.4in]{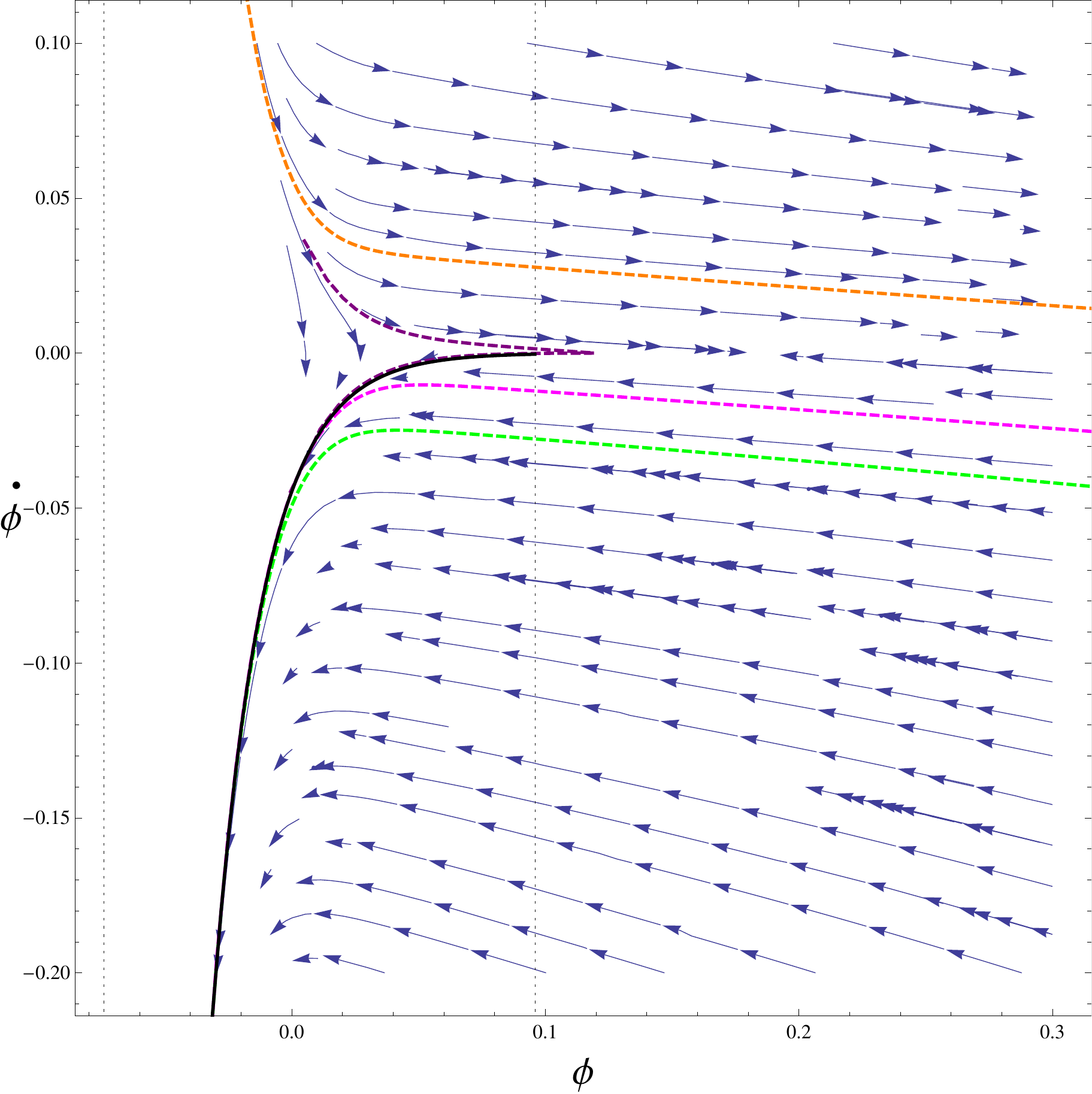}}
   \subfigure[Contracting]{\includegraphics[width=3.4in]{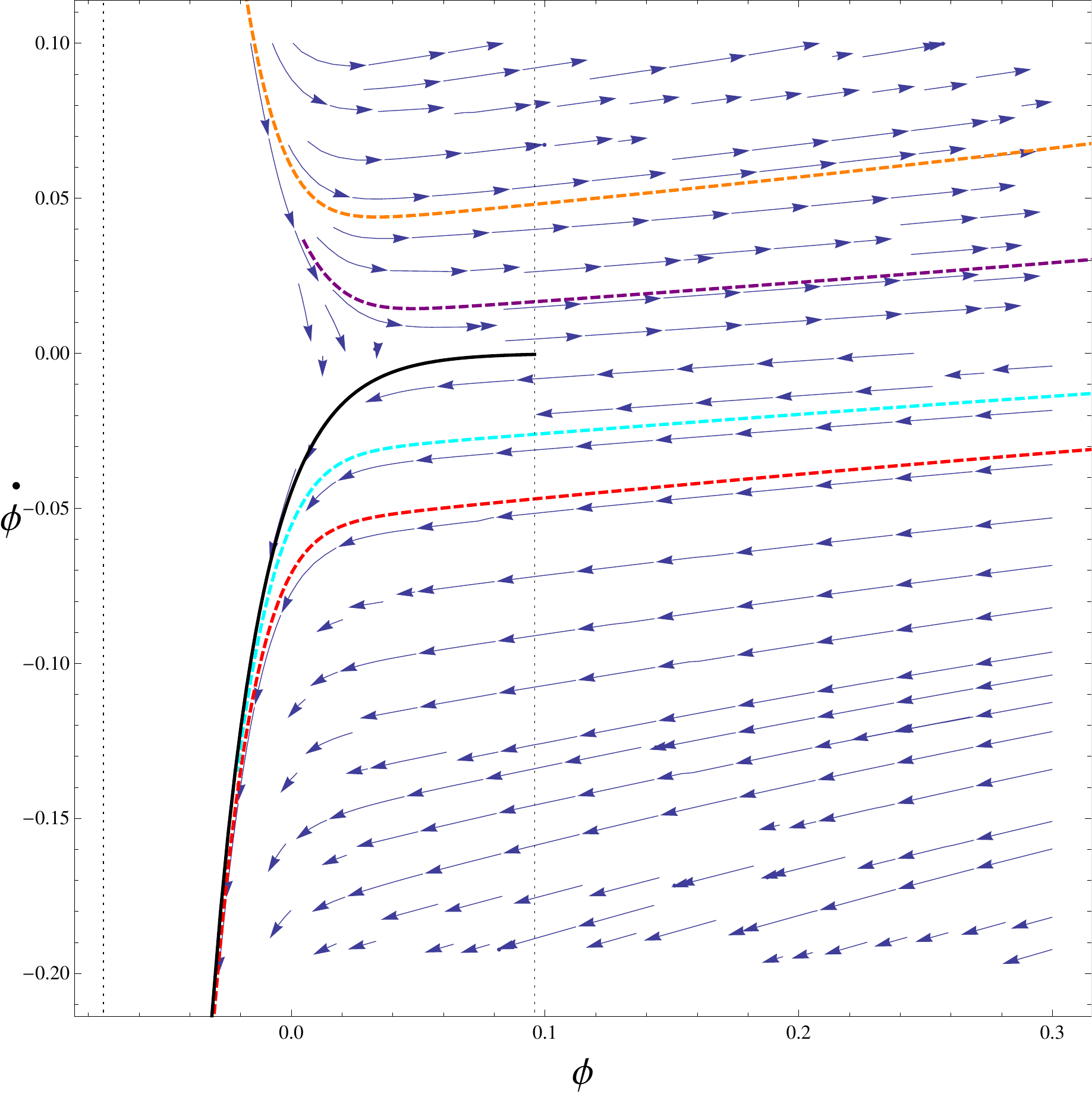}}
   \caption{Phase portrait for the expanding (a) and contracting (b) cases, for $c=100$ and $V_0=10^{-4}$. The analytic solution is denoted by the black curve, with the transition phase taking place between the dotted lines. Colored dashed lines are particular numerical solutions to the system~(\ref{phidotphivector}). Note that the arrows point in the direction of increasing time. This Figure confirms that the analytic solution is an attractor for a variety of initial conditions in both cases.  However, at large positive $\phi$ where $V(\phi)\approx V_0$, the expanding solution is also an attractor, while the contracting solution is a repellor, due to the asymmetry between expanding vs. contracting de Sitter space.}
 \label{globalview}
\end{figure}
\begin{figure}
   \centering
   \subfigure[Expanding]{\includegraphics[width=3.4in]{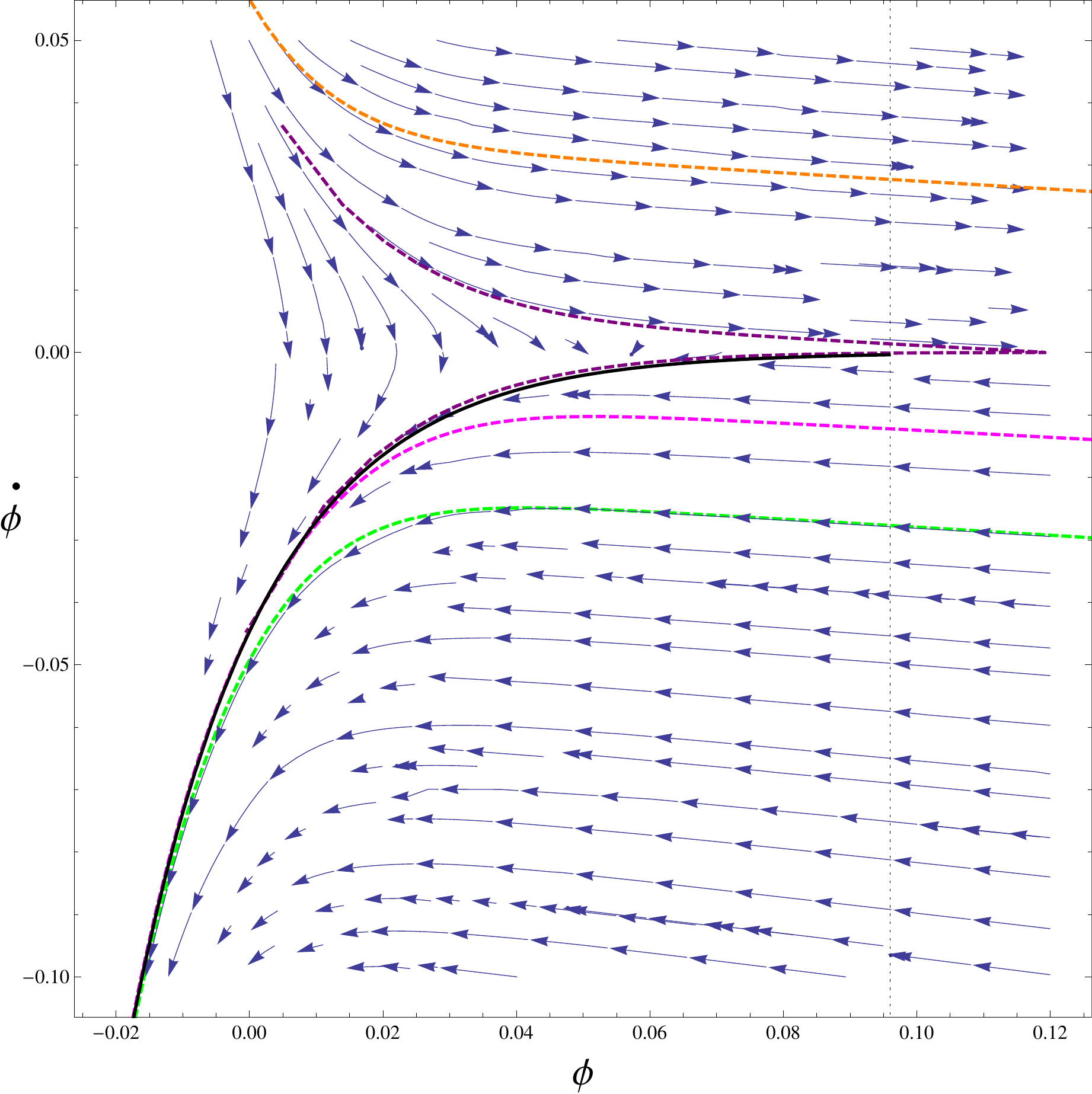}}
   \subfigure[Contracting]{\includegraphics[width=3.4in]{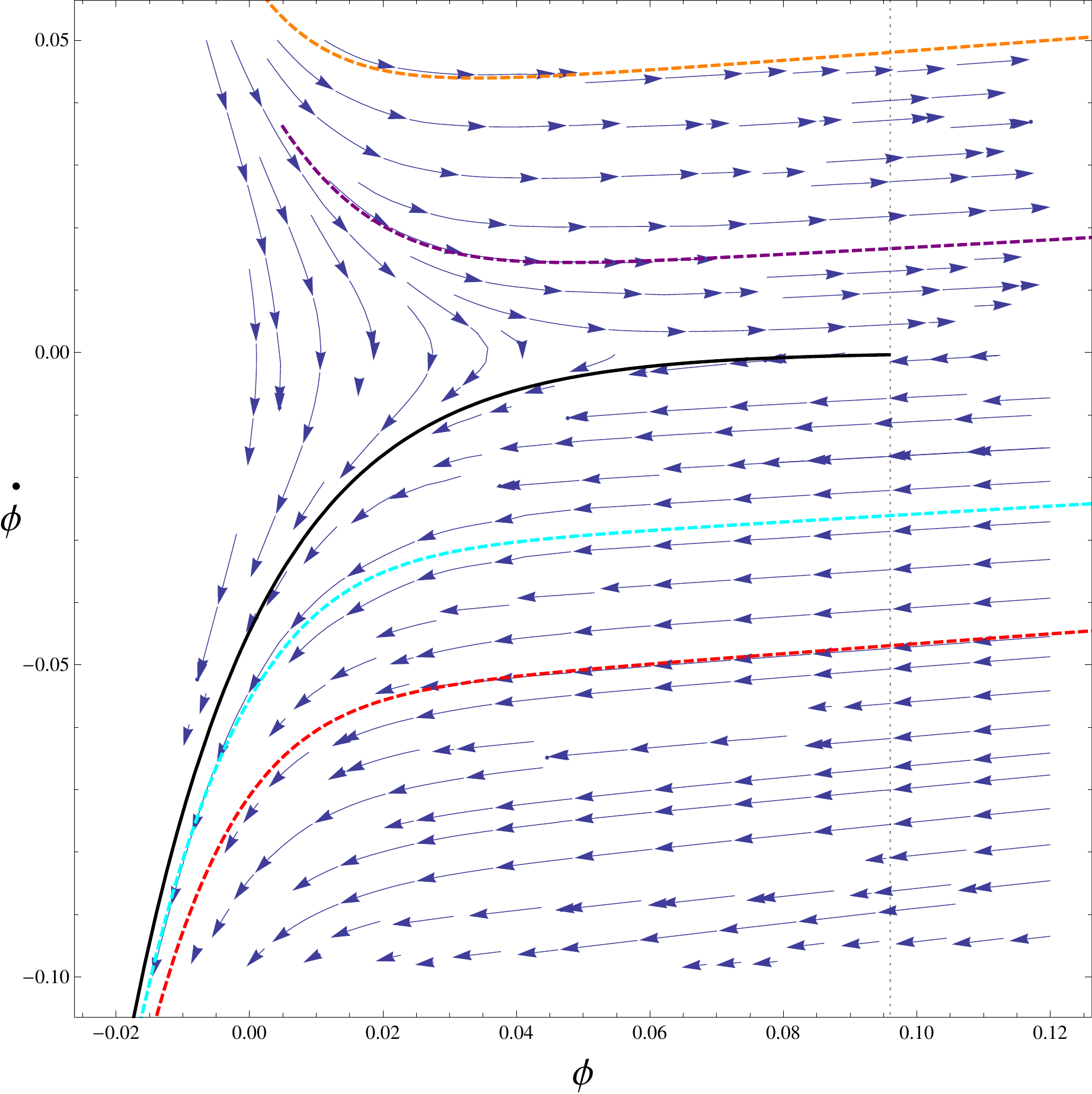}}
   \caption{Detailed view of Fig.~\ref{globalview}, zooming in on the transition phase.
   Both contracting and expanding transition solutions are attractors, but the expanding case has a slightly larger basin of attraction.}
\label{transitionview}
\end{figure}

\subsection{Numerical Results}

Figure~\ref{globalview} shows the vector field $({\rm d}\phi/{\rm d}N, {\rm d}\dot{\phi}/{\rm d}N)$ given by~(\ref{phidotphivector}) for the expanding and
contracting cases, respectively, along with some numerically integrated curves. The black curve in each case corresponds to the analytic
solution, with the transition phase occurring between the dotted lines. The chosen parameters are $c=100$ and $V_0=10^{-4}$. Figure~\ref{transitionview} zooms in on the transition phase. 

As is clear from Fig.~\ref{transitionview}, the transition solution is an attractor, both in the expanding and contracting cases. This confirms our earlier perturbative claims based on
the long-wavelength behavior of $\zeta$. The expanding solution has a slightly larger basin of attraction, as expected. In particular, note that in the contracting case curves with $\dot\phi$ sufficiently greater than zero are forced away from the transition solution while in the expanding case, such curves are driven toward the slow-roll solution and then follow the transition solution.

More globally, Fig.~\ref{globalview} shows that in the contracting case curves are repelled from the `slow-roll' solution at large $\phi$. This is due to the susceptibility of contracting de Sitter to kinetic domination. In the expanding case however, we see the opposite behavior --- trajectories with significant initial kinetic energy are driven toward the slow-roll de Sitter solution before undergoing the transition phase evolution.
To summarize, both the expanding and contracting transition solutions are attractors for some range of initial conditions, but the expanding solution has a larger basin of attraction.

\section{Fusion of Slow-Roll and Transition Solutions}
\label{fusion}

One of our motivations for considering an initially expanding cosmology was the instability of the contracting solution at sufficiently early times.
The adiabatic ekpyrotic mechanism requires that the transition phase starts out with a sufficiently small equation of state parameter, $\epsilon \ll 1$,
which implies that prior evolution will generically be unstable to kinetic domination in the contracting case.

As shown in Sec.~\ref{phasespace}, this is true, in particular, for the simplest potential~(\ref{scalarpotential}) --- the pre-transition phase evolution corresponds to a contracting de Sitter universe,
which is clearly unstable. The expanding branch, on the other hand, is better behaved, as it extrapolates backwards in time to an expanding de Sitter space.
Thus, in choosing an initially expanding universe, we have greatly expanded the basis of attraction. (It is worth emphasizing that even in the initially expanding case, the universe is 
driven towards a contracting ekpyrotic scaling phase.)

As mentioned in the Introduction, our fiducial potential~(\ref{scalarpotential}) serves as the simplest illustration of our mechanism. More generally, this form need only hold
approximately during the transition phase, corresponding to a small range in field space, $\Delta\phi \ll M_{\rm Pl}$, and there is ample freedom in choosing the
global form of the potential. That said, if we take the potential~(\ref{scalarpotential}) at face value, then at sufficiently early times the evolution in the expanding case
should correspond to an inflating space-time. In this Section, we take this possibility seriously and study the transition between the initial de Sitter phase to the transition phase.
We will see that $\phi(t)$ evolves smoothly between the two regimes.

To see this, let us split the evolution into two regimes:
\newcounter{Lcount}
\begin{list}{\Roman{Lcount}}
{\usecounter{Lcount}}
\item The `transition' regime, $\phi<\phi_{\rm T}$, where Hubble friction is negligible and the scalar field equation of motion reduces to~(\ref{scalareomapprox}):
\be
\ddot\phi+V_{,\phi}=0\ .
\label{eomtran}
\ee
\item The `slow-roll' regime, $\phi > \phi_{\rm T}$, where the equation of motion is approximately given by
\be
3H\dot\phi+V_{,\phi}=0\ ,
\label{eomslow}
\ee
with $H \simeq H_0 = \sqrt{V_0/3M_{\rm Pl}^2}$.
\end{list}
The field value delineating these two regions, $\phi_{\rm T}$, can be estimated as the value of $\phi$ at the onset of the transition phase. Combining~(\ref{phisol2}),~(\ref{H0}) and~(\ref{tbeg}),
we obtain
\be
\phi_{\rm T} = \phi(t_{\rm beg-tran}) = \frac{2M_{\rm Pl}}{c}\log \left(\sqrt{\frac{3}{2}}c\right)\simeq \frac{2M_{\rm Pl}}{c}\log c\,,
\ee
where in the last step we have used $c\gg 1$.

We are now in a position to argue that the approximate solutions in regions I and II match each other smoothly at $\phi = \phi_{\rm T}$. Since the potential is certainly continuous,
this amounts to showing that the kinetic energy also matches continuously. The kinetic energy of the transition solution is easy to write down, as we have done it several times
already, and is just the first integral of~(\ref{eomtran}):
\be
\frac{1}{2}\dot\phi^2_{\rm tran}\approx V_0e^{-c\phi/M_{\rm Pl}}\ .
\label{KEtran}
\ee
Likewise, the kinetic energy in the slow-roll regime follows from~(\ref{eomslow}):
\be
\frac{1}{2}\dot\phi^2_{\rm slow-roll}\approx\frac{V_{,\phi}^2M_{\rm Pl}^2}{6V_0}\approx V_0c^2e^{-2c\phi/M_{\rm Pl}} \,.
\label{KEsr}
\ee
It immediately follows that~(\ref{KEtran}) and~(\ref{KEsr}) are equal at $\phi = \phi_{\rm T}$, as we wanted to show.
This result tells us that trajectories starting in the slow roll region, $\phi > \phi_T$, are quickly driven to the attractor
slow-roll solution and reach $\phi_T$ with precisely the correct kinetic energy as assumed by the transition solution.
(In the notation of Sec.~\ref{rest}, solutions that emerge from the slow regime generically reach the transition phase
with $E \simeq V_0$.)

Note that the spectrum of fluctuations also matches smoothly between the two regimes. The power spectrum of modes generated during the inflationary epoch is given by
\be
P_\zeta^{\rm inf} =\frac{1}{(2\pi)^2}\frac{H^4}{\dot{\phi}^2_{\rm slow-roll}} \simeq\frac{c^2V_0}{12\pi^2}\,,
\ee
which agrees up to an order unity factor with~(\ref{power}).

\section{Conclusions}
\label{conclu}

The adiabatic ekpyrotic mechanism is the unique, non-inflationary single-field mechanism with unit sound speed that generates a scale invariant
spectrum for $\zeta$ on an attractor background. While originally proposed assuming a contracting universe, in this paper
we have shown that the mechanism works equally well on an initially expanding background. The evolution consists of
an ``expanding transition phase", followed by a contracting ekpyrotic scaling phase. 

We have shown, both through analytical arguments and exact numerical integration, that the power spectrum for $\zeta$ is indistinguishable from
the contracting version of the mechanism. This confirms that the perturbation spectrum is to a good approximation insensitive to whether the background
is expanding or contracting during mode production --- scale invariance relies on a rapidly-varying equation of state parameter while the scale factor is nearly constant.

By performing a phase space analysis for both the expanding and contracting branches, we have verified that the transition phase and subsequent ekpyrotic scaling phase are attractors. The basin of attraction is broader in the expanding case, since intuitively any additional kinetic energy present at early times gets redshifted, instead of blueshifted,
in this case.

For the simplest potentials considered here, the evolution is an expanding de Sitter space-time asymptotically in the past. Taking this precursor inflating phase seriously,
we have shown that the transition from inflation to the adiabatic ekpyrotic phase is smooth, with the scale invariant spectra generated in each phase matching at the transition, up to
order unity coefficient.

As mentioned in the Introduction, the degeneracy with inflation is broken at the three-point level. Unlike the nearly Gaussian spectrum of inflation, the
rapidly-varying equation of state $\epsilon\sim 1/\tau^2$ characteristic of the adiabatic ekpyrotic phase leads to large non-Gaussianities on small scales. 
For the simplest lifted exponential potential~(\ref{scalarpotential}), this results in a breakdown of perturbation theory, both at the classical and quantum
mechanical levels. As shown in detail in~\cite{Khoury:2011ii} in the contracting case, however, this perturbative breakdown can be avoided for more general potentials of the form~(\ref{genpot}),
but this comes at a cost --- the range of scale invariant perturbations is now limited, spanning a factor of $10^5$ in $k$ space. These considerations should carry over to the expanding case.
We leave a careful exploration of this issue to future work.

The expanding version of adiabatic ekpyrosis presented here greatly expands the realm of larger cosmological scenarios in which this
mechanism can be embedded. Because the basin of attraction is broader compared to its contracting counterpart, the expanding mechanism is
less sensitive to the details of the prior evolution. More importantly for model-building, the universe is expanding during mode production
and is contracting later on, which naturally suggests embedding our mechanism in a cyclic scenario~\cite{Steinhardt:2001st}. It is conceivable that tying
our mechanism to the present phase of cosmic acceleration may also explain why the $10^5$ scale invariant modes fall in the observable window.
We are currently investigating this possibility.

\textit{Acknowledgments:} It is our pleasure to thank Daniel Baumann, Adam Brown, Kurt Hinterbichler, Mitchell Lerner, Godfrey Miller, Leonardo Senatore, Paul Steinhardt, Daniel Wesley and Matias Zaldarriaga for helpful discussions. This work was supported in part by funds from the University of Pennsylvania, the Alfred P. Sloan Foundation (JK) and the Ella N. Pawling Fellowship from the University of Pennsylvania (AJ).

\end{document}